\documentclass[prl,aps,twocolumn,showpacs]{revtex4}
\usepackage{graphicx}
\usepackage{amsmath}
\usepackage{bm}
\usepackage{amstext}
\usepackage{amsxtra}
\usepackage{gensymb}

\usepackage{times}

\begin{document}

\newcommand{\To}{T_c^0}
\newcommand{\kB}{k_{\rm B}}
\newcommand{\dT}{\Delta T_c}
\newcommand{\lo}{\lambda_0}
\newcommand{\cs}{$\clubsuit$}
\newcommand{\thold}{t_{\rm hold}}
\newcommand{\Nmf}{N_c^{\rm MF}}
\newcommand{\Neq}{N_0^{\rm eq}}
\newcommand{\Tmf}{T_c^{\rm MF}}
\newcommand{\el}{\gamma_{\rm el}}
\newcommand{\meq}{\mu^{\rm eq}}
\newcommand{\n}{\tilde{n}}
\newcommand{\tA}{\theta_{\alpha}}

\title{Bose-Einstein condensation of atoms in a uniform potential}

\author{Alexander L. Gaunt, Tobias F. Schmidutz, Igor Gotlibovych, Robert P. Smith, and Zoran Hadzibabic}
\affiliation{Cavendish Laboratory, University of Cambridge, J.~J.~Thomson Ave., Cambridge CB3~0HE, United Kingdom}

\begin{abstract}
We have observed Bose-Einstein condensation of an atomic gas in the (quasi-)uniform three-dimensional potential of an optical box trap. Condensation is seen in the bimodal momentum distribution and the anisotropic time-of-flight expansion of the condensate. The critical temperature agrees with the theoretical prediction for a uniform Bose gas. The momentum distribution of our non-condensed quantum-degenerate gas is also clearly distinct from the conventional case of a harmonically trapped sample and close to the expected distribution in a uniform system. 
We confirm the coherence of our condensate in a matter-wave interference experiment.
Our experiments open many new possibilities for fundamental studies of many-body physics.
\end{abstract}

\date{\today}



\maketitle

Ultracold Bose and Fermi atomic gases are widely and successfully used as testbeds of fundamental many-body physics \cite{Bloch:2008}. 
Experimental tools such as Feshbach interaction resonances~\cite{Chin:2010}, optical lattices~\cite{Morsch:2006}, and synthetic gauge fields~\cite{Dalibard:2012}, offer great flexibility for studies of outstanding problems arising in many areas, most commonly in condensed matter physics.
However, an important difference between ``conventional" many-body systems and ultracold gases is that the former are usually spatially uniform while the latter are traditionally produced in harmonic traps with no translational symmetries.

Various methods have been developed to overcome this problem and extract uniform-system properties from a harmonically trapped sample \cite{Ho:2010a, Nascimbene:2010, Navon:2010, Hung:2011, Yefsah:2011, Ku:2012, Smith:2011b,  Drake:2012,Sagi:2012}, relying on the local density approximation \cite{Ho:2010a, Nascimbene:2010, Navon:2010, Hung:2011, Yefsah:2011, Ku:2012, Smith:2011b} or selective probing of a small central portion of the cloud \cite{Smith:2011b,Drake:2012,Sagi:2012}. 
Sometimes harmonic trapping can even be advantageous, allowing simultaneous mapping of uniform-system properties at different (local) particle densities. On the other hand, in many important situations local approaches are inherently limiting, for example for studies of critical behaviour with diverging correlation lengths. The possibility to directly study a spatially uniform quantum-degenerate gas has thus remained an important experimental challenge.
So far, atomic Bose-Einstein condensates (BECs) have been loaded into linear \cite{Meyrath:2005} or circular \cite{Gupta:2005} traps which are uniform along only one direction while still being harmonic along the other two. 

\begin{figure} [bp]
\includegraphics[width=\columnwidth]{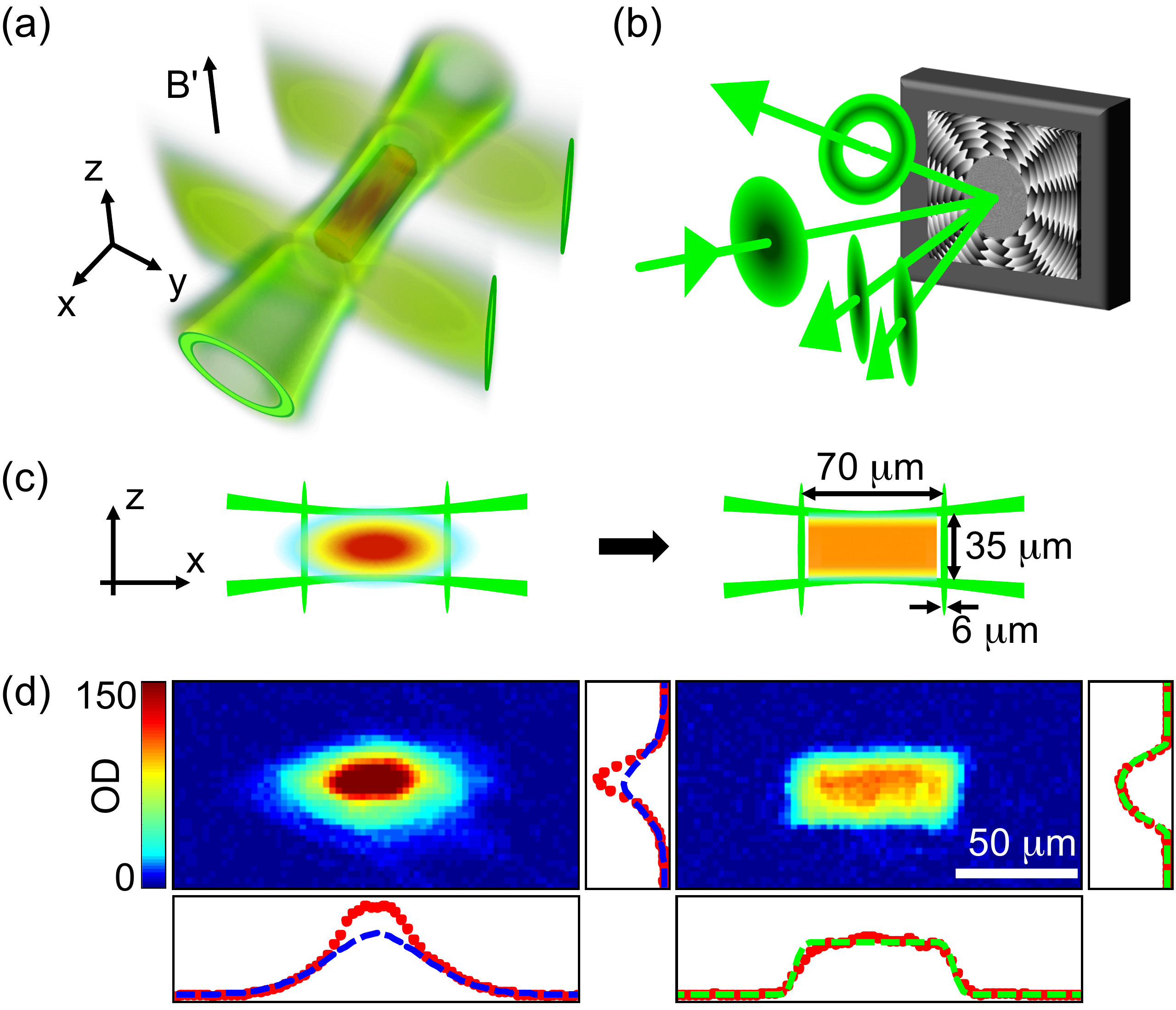}
\caption{Preparing a quasi-uniform Bose gas. (a) The optical-box trap is formed by one hollow ``tube" beam and two ``sheet" beams creating a repulsive potential for the atoms. The atomic cloud, depicted in red, is confined to the dark cylindrical region. Gravitational force is cancelled by a magnetic field gradient $B'$. (b) The three trapping beams are created by reflecting a single Gaussian beam off a phase-imprinting spatial light modulator. (c) The atoms are loaded into the box trap after pre-cooling in a harmonic trap. 
(d) {\it In situ} images of the cloud just before (left) and after (right) loading into the box, and corresponding line-density profiles along $x$ (bottom plots) and $z$ (side plots). OD stands for optical density. The blue dashed lines are fits to the thermal component of the harmonically trapped gas.  The green dashed lines are fits based on the expected profiles for a uniform-density gas.}
\label{fig:box}
\end{figure}

Here we demonstrate Bose-Einstein condensation of an atomic gas in a three-dimensional (quasi-)uniform potential. We load an optical box trap depicted in Fig.~\ref{fig:box}(a) with $^{87}$Rb atoms pre-cooled in a harmonic trap and achieve condensation by evaporative cooling in the box potential.  Below a critical temperature, $T_c \approx 90\;$nK,
condensation is seen in the emergence of a bimodal momentum distribution and the anisotropic time-of-flight (TOF) expansion of the BEC. We characterise the flatness of our box potential and show that both the momentum distribution of the non-condensed component and the thermodynamics of condensation are close to the theoretical expectations for  a uniform system, while being clearly distinct from the conventional case of a harmonically trapped gas. We also demonstrate a simple experimental configuration for trapped-atom interferometry and use it to confirm the coherence of our quasi-uniform BEC.  

Our setup for producing $^{87}$Rb condensates in a harmonic potential is described elsewhere \cite{Gotlibovych:2013}; we create BECs in the $|F, m_F\rangle = |2,2\rangle$ hyperfine ground state using a hybrid magnetic-optical trap~\cite{Lin:2009}.
The dark optical trap \cite{Kaplan:2002, Jaouadi:2010} which is central to this work is formed by three 532~nm laser beams - a ``tube" beam propagating along the $x$ axis and two ``sheet" beams propagating along $y$. 
The green laser beams create a repulsive potential for the atoms and confine them to the cylindrical dark region depicted in red in Fig.~\ref{fig:box}(a).
To create a uniform potential, we additionally cancel the gravitational force on the atoms at a $10^{-4}$ level, using a magnetic field gradient~\cite{Gotlibovych:2013}.

As outlined in Fig.~\ref{fig:box}(b), all three trapping beams are created by reflecting a single Gaussian beam off a phase-imprinting spatial light modulator (SLM) with three superposed phase patterns \cite{Liesener:2000}. The tube beam is an optical vortex created by imprinting a $24\, \pi$ phase winding on the incoming beam, the sheet beams are created using cylindrical-lens phase patterns, and the three outgoing beams are deflected in different directions using phase gradients.
With a total laser power of $P_0 \approx 700\;$mW we achieve a trap depth of $V_0 \approx \kB \times 2\;\mu$K.

We evaporatively cool the gas in the harmonic trap down to $T \approx 120$~nK, when the cloud size is similar to the size of our optical box [see Fig.~\ref{fig:box}(c)] and $\kB T \ll V_0$. At this point the gas is partially condensed, but the BEC is lost during the transfer into the box trap, which is not perfectly adiabatic. Over 1~s, we turn on the green light and then turn off the harmonic trapping, capturing $> 80\,\%$ of the atoms. 

In Fig.~\ref{fig:box}(d) we show in-trap absorption images of the cloud just before and just after the transfer into the box trap.
The images are taken along the $y$ direction, using high intensity imaging \cite{Reinaudi:2007,Hung:2011,Yefsah:2011} with a saturation parameter $I/I_{\rm sat} \approx 150$. For each image we show the line-density profiles along $x$ and $z$.
If a cylindrical box of length $L$ and radius $R$ is filled perfectly uniformly, the density distribution along $x$ is simply a top-hat function of width $L$. Along $z$, the line-of-sight integration results in ``circular" column- and line-density profiles, $\propto \sqrt{1 - (z/R)^2}$.
In the experimental images, the edges of the cloud are rounded off for two reasons, both related to the diffraction limit of our optical setup. First, the waist of the $532\;$nm trapping beams is diffraction limited to $\approx 3\;\mu$m, which leads to some rounding-off of the potential bottom near the edges of the box. Second, our imaging resolution is diffraction limited to $\approx 5\;\mu$m, making the cloud edges appear more smeared out than they actually are.
The green dashed lines in Fig.~\ref{fig:box}(d) are fits to the data based on a perfectly uniform distribution convolved {\it only} with the imaging point-spread function. The fits describe the data well and give $L=63 \pm 2\;\mu$m and $R=15 \pm 1\;\mu$m. These values are consistent with the calculated separation of the green walls, reduced by the diffraction-limited wall thickness.

After the transfer into the box trap the cloud contains $N \approx 6 \times 10^5$ atoms at $T \approx 130\;$nK.
From this point, we cool the gas to below $T_c$ by forced evaporative cooling in the box trap, lowering the trapping power $P$ in an exponential ramp with a 0.5~s time constant. Initially the trap depth is much larger than $\kB T$ so significant cooling occurs only for $P \lesssim 0.5\,P_0$.

\begin{figure} [tbp]
\includegraphics[width=\columnwidth]{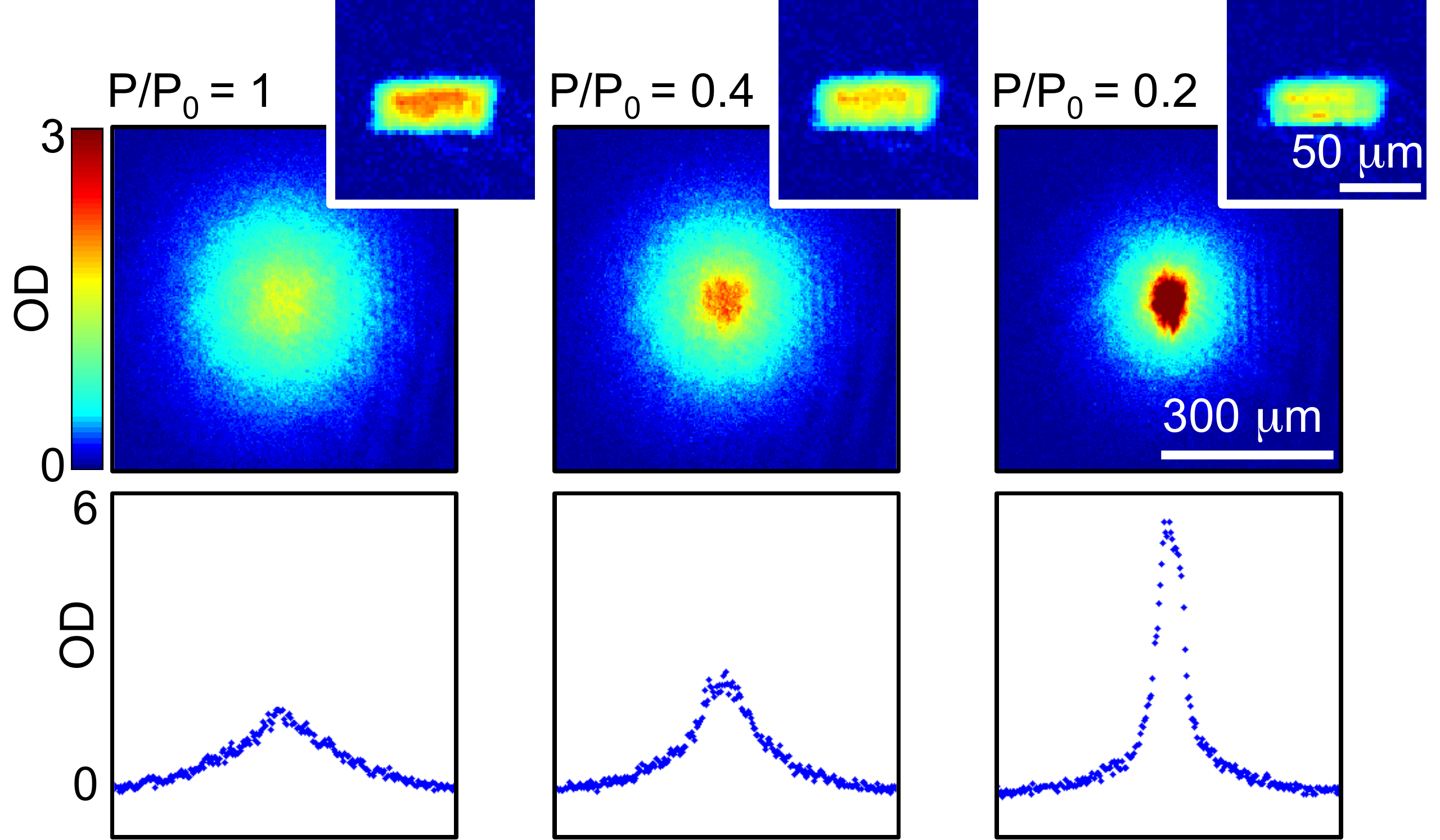}
\caption{Evaporation and Bose-Einstein condensation in the optical-box trap. Cooling is achieved by lowering the trapping laser power, $P$.
We show absorption images taken after 50 ms of TOF and {\it in situ} [insets, with same colour scale as in Fig.~\ref{fig:box}(d)]. The bottom panels show cuts through the momentum distributions recorded in TOF. 
In contrast to the case of a harmonic trap, no dramatic effects of cooling are observed {\it in situ}.   
However, BEC is clearly seen in the bimodality of the momentum distribution and the anisotropic expansion.}
\label{fig:BEC}
\end{figure}

Fig.~\ref{fig:BEC} qualitatively illustrates the effects of evaporation and condensation in the box trap. 
We show images of the cloud both {\it in situ} and after $t=50$~ms of TOF expansion from the trap. 
While in a harmonic trap cooling results in simultaneous real-space and momentum-space condensation, here it has no dramatic effects on the in-trap atomic distribution. The density is gradually reduced by evaporation, but the shape of the cloud does not reveal condensation. On the other hand, in momentum space (i.e., in TOF) the effects of cooling are obvious and the signatures of condensation are qualitatively the same as for a harmonically trapped gas - the momentum distribution becomes bimodal and the BEC expands anisotropically, with its aspect ratio inverting in TOF. 

We now turn to a quantitative analysis of our degenerate quasi-uniform Bose gas. We assess the flatness of our trapping potential and contrast the thermodynamics of condensation in our system with the case of a harmonically trapped gas.

\begin{figure*} [tbp]
\includegraphics[width=\textwidth]{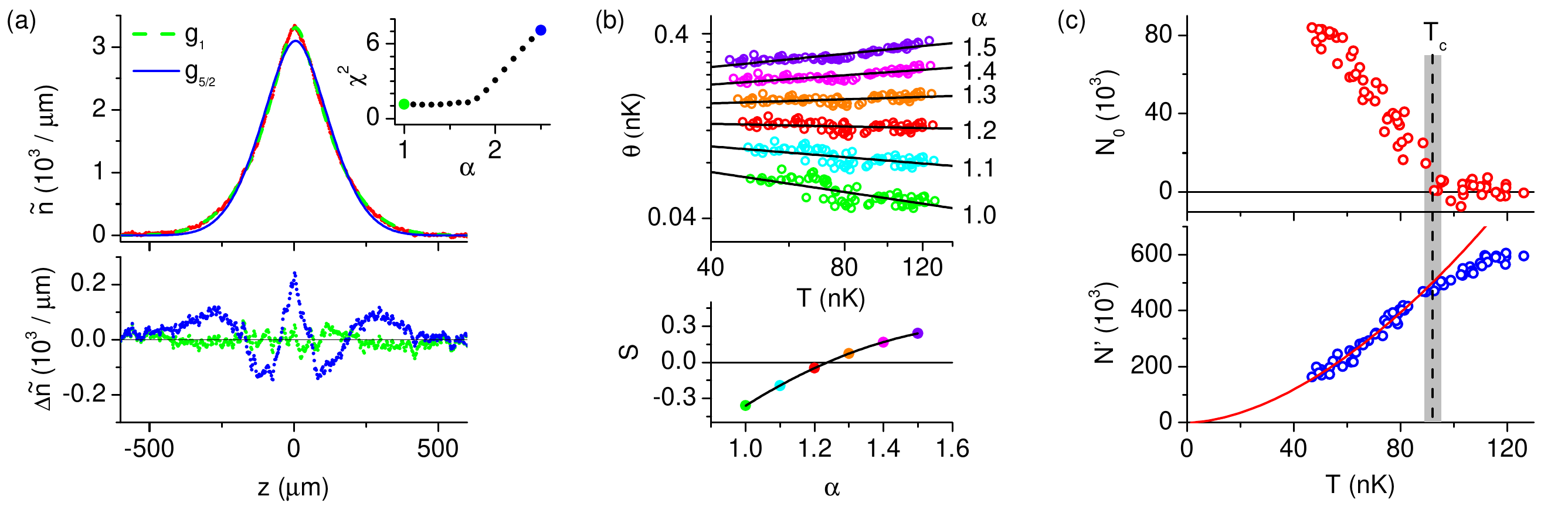}
\caption{Momentum and particle-number distribution in a quasi-uniform degenerate Bose gas. 
(a) TOF profile $\n(z)$ of a non-condensed cloud (red) and $g_{\alpha}$ fits to the data, with  $\alpha =1$ (green) and $5/2$ (blue). We also show the fitting residuals, $\Delta \n$. Inset: fitting $\chi^2$ versus $\alpha$.
(b) Flatness of the box potential: the condition $S = {\rm d} \left[\log(\theta)\right]/{\rm d}[\log(T)] =0$ (see text) gives $\alpha = 1.23 \pm 0.03$, corresponding to an effective $r^{13\pm 2}$ potential.
(c) The number of condensed ($N_0$) and thermal ($N'$) atoms versus $T$. The critical temperature, $T_c = 92 \pm 3\;$nK, is in agreement with the prediction for a uniform Bose gas. Below $T_c$, a power-law fit (solid red line) gives $N' \propto T^{1.73 \pm 0.06}$.
}
\label{fig:plots}
\end{figure*}

We first analyse the momentum distribution in a non-condensed gas, at $T \approx 110\;$nK. In Fig.~\ref{fig:plots}(a) we plot the line density measured in TOF, $\n(z)$, obtained by integrating an image along $x$. At this point the gas is sufficiently degenerate to show the effects of Bose statistics. In a trapped degenerate gas the spatial and momentum degrees of freedom are in general not separable and consequently $\n(z)$ also contains information about the functional form of the trap. 

For simplicity let us consider an isotropic 3D trapping potential of the form $V(r) \propto r^n$; for a fixed $n$, potential anisotropy does not affect the scalings discussed below.
The line-density distribution after long TOF then has the form
\begin{equation}
\n(z) = \sqrt{\frac{\beta m}{2t^2}} \left( \frac{T}{\tA} \right)^{\alpha+1/2}  g_{\alpha} \left( e^{\beta \left( \mu - \varepsilon(z) \right)}  \right) \, ,
\end{equation}
where $g_{\alpha}$ is the polylog function of order $\alpha = 1 + 3/n$. Here $\beta = 1/(\kB T)$, $m$ is the atom mass, $\mu$ is the chemical potential, $\varepsilon (z) = mz^2/(2t^2)$, and the constant $\tA$ absorbs various factors such as the imaging magnification and cross section. Small corrections due to the initial cloud size are accounted for by convolving $g_{\alpha}$ with the in-trap density distribution.

For a harmonic trap $n=2$ and $\alpha = 5/2$, while for a uniform gas $n \rightarrow \infty$ and $\alpha = 1$. The measured $\n$ is fitted very well by the $g_1$ function, with the reduced $\chi^2 \approx 1$.  The  $g_{5/2}$ fit is comparatively poor: $\chi^2 \approx 7$ and the systematic patterns in the  fitting residuals, $\Delta \n = \n - \n_{\rm fit}$, clearly show that this is fundamentally a wrong functional form. Qualitatively, the measured momentum distribution is more ``peaky" than that of a harmonically trapped degenerate gas. 

As shown in the inset of Fig.~\ref{fig:plots}(a), the measured $\n$ can be fitted well using $\alpha$ in the range $1 - 1.7$, while higher values can be clearly excluded. Note however that in these fits we used $\tA$, $\mu$ and $T$ as free parameters. 
Allowing $\tA$ to vary gives the fitting function an unphysical freedom and overestimates the range of suitable $\alpha$ values. 
Crucially, {\it only} for the correct $\alpha$ is the best-fit value of $\tA$ a temperature-independent constant. 
We use this fact to accurately determine the leading-order correction to the flatness of our trapping potential.

In Fig.~\ref{fig:plots}(b) we analyse the $T$-dependence of the fitted $\tA$ values. 
Here, we exclude from the fits the central $180\; \mu$m wide region of the cloud, which is larger than the largest observed BECs.
The drift of the fitted $\tA$  with $T$ is described well by a constant slope $S_{\alpha} = {\rm d} \left[\log(\tA)\right]/{\rm d}[\log(T)]$, and we see that $S_{\alpha}$ varies monotonically with $\alpha$.
Physically, if the $\alpha$ used for fitting is too high, as the gas cools the actual momentum distribution narrows faster than $g_{\alpha}$. The fit then increasingly underestimates $T$ and compensates by decreasing $\tA$. 
Conversely, if $\alpha$ is too low the fitted $\tA$ increases as the gas is cooled.
Note that for each $\alpha$ the fitted $\tA$ is plotted against $T$ extracted from the same fit, and we see that the fitted temperatures also show the expected systematic drifts.

From the condition $S_{\alpha} = 0$ we get $\alpha = 1.23 \pm 0.03$ and conclude that the leading-order correction to the flatness of our box potential is $ \propto r^{13\pm2}$.
While we can distinguish an $r^{13}$ potential from a perfectly flat one, we expect this distinction to be irrelevant for most many-body studies.

We now fix $\alpha =1.23$ and study the evolution of the BEC atom number, $N_0$, with $T$. The cylindrical BEC does not follow any simple scaling in TOF, 
so rather than fitting it to any specific shape we fit only the thermal component and simply count the excess atoms not accounted for by the thermal fit.

In Fig.~\ref{fig:plots}(c) we plot $N_0$ versus $T$ and determine $T_c = 92 \pm 3\;$nK.
From the measured atom number and box volume, we get a consistent theoretical value for a uniform Bose gas, $\To = 98 \pm 10\;$nK.
This calculation includes small finite-size and interaction shifts of $\To$  \cite{Grossmann:1995,Arnold:2001,Kashurnikov:2001}, about $6\,\%$  and $1\,\%$ respectively; 
the error includes the $10\,\%$ uncertainty in our absolute atom-number calibration, obtained by measuring $T_c$ in a harmonic trap \cite{Smith:2011}.

In Fig.~\ref{fig:plots}(c) we also plot the thermal atom number, $N'$, versus $T$.  Below $T_c$, we fit the variation of $N'$ with a power law, $N' \propto T^{\gamma}$, and get $\gamma = 1.73 \pm 0.06$.
This agrees with the expected $\gamma = \alpha +1/2$.
For a saturated thermal component in a perfectly uniform system  $N' \propto T^{3/2}$, while in a harmonic trap $N' \propto T^3$. 
The thermodynamics of our gas are therefore very close to the textbook case of a uniform system and very different from the case of a harmonically trapped sample.

Below $40\;$nK our temperature fits are not accurate, but we can cool the gas further and produce a quasi-pure BEC with $>10^5$ atoms. The $1/e$ lifetime of the BEC is $10\;$s, limited by the background pressure in our vacuum chamber \cite{Gotlibovych:2013}. 

\begin{figure} [tbp]
\includegraphics[width=\columnwidth]{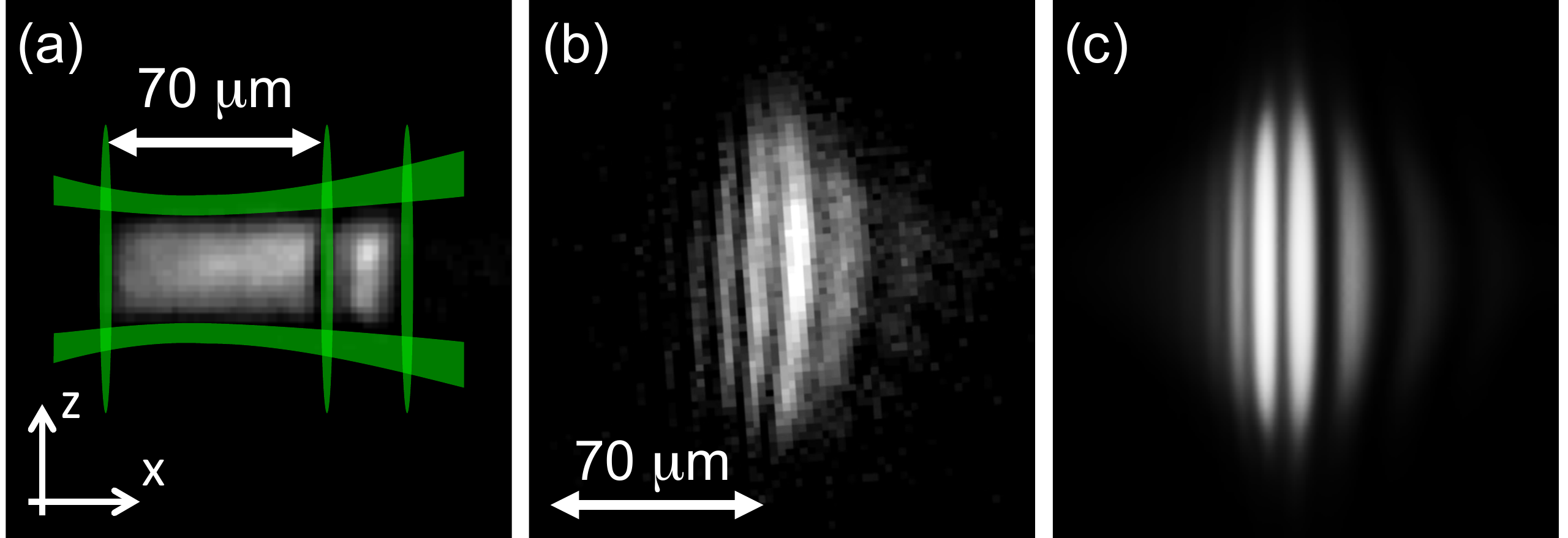}
\caption{Matter-wave interference with a quasi-uniform BEC. (a) {\it In situ} image of a two-trap configuration, together with a sketch of the green-light arrangement used to create it.  (b) Interference pattern observed after 100 ms of TOF expansion. (c) Numerical simulation of the expected interference pattern.}
\label{fig:interference}
\end{figure}

Finally, we verify the coherence of our uniform BEC in an interference experiment, using a two-trap configuration shown in Fig.~\ref{fig:interference}(a). 
We add a third ``sheet" laser beam to simultaneously condense atoms in a small ``satellite" trap, while the main box trap is essentially the same as in the rest of the paper. 

In Fig.~\ref{fig:interference}(b) we show the interference fringes emerging as the two clouds overlap in TOF. Note that the interference pattern is different from the parallel fringes of constant spacing  commonly observed in a symmetric expansion of two BECs \cite{Andrews:1997a}. Qualitatively, along $x$ the main cloud expands very slowly and is still in the ``near-field" regime, while the small BEC expands rapidly and is in the far-field regime. This leads to a quadratic variation of the relative phase along $x$ and the reduction of the fringe period from right to left in Fig.~\ref{fig:interference}(b). We reproduce a similar pattern in a numerical simulation of the Gross-Pitaevskii equation, shown in Fig.~\ref{fig:interference}(c).

In conclusion, we have observed and characterised Bose-Einstein condensation of an atomic gas in an essentially uniform potential.
We expect our experiments to open up many new research possibilities.
The box trap eliminates the need to rely on the local density approximation, which could be particularly important for studies of critical behaviour with diverging correlations near phase transitions. 
Moreover, for the same particle number, the quasi-uniform BEC has a significantly lower density (in our case $\sim 2 \times 10^{12}\;$cm$^{-3}$) than a harmonically trapped one. This
reduces the importance of three-body recombination compared to two-body interactions and near a Feshbach resonance it may facilitate equilibrium studies of a strongly interacting, possibly unitary Bose gas \cite{Navon:2011,Wild:2012,Li:2012}. 
Dark optical traps loaded with low density BECs also hold promise for trapped-atom interferometry with very long decoherence times.
Here we chose an elongated box shape for qualitative detection of the BEC through anisotropic expansion, but with our light-shaping methods traps of almost arbitrary geometry could be created~\cite{Gaunt:2012}.
Our methods are also suitable for studies of degenerate Fermi gases and are compatible with implementation of 3D optical lattices. 

This work was supported by EPSRC (Grant No. EP/G026823/1), the Royal Society, AFOSR, ARO and DARPA OLE. We thank Laser Quantum for the loan of the 532 nm laser.


\begin{thebibliography}{10}

\bibitem{Bloch:2008}
I. Bloch, J. Dalibard, and W. Zwerger, Rev. Mod. Phys. {\bf 80},  885  (2008).

\bibitem{Chin:2010}
C. Chin, R. Grimm, P. Julienne, and E. Tiesinga, Rev. Mod. Phys. {\bf 82},
  1225  (2010).

\bibitem{Morsch:2006}
O. Morsch and M. Oberthaler, Rev. Mod. Phys. {\bf 78},  179  (2006).

\bibitem{Dalibard:2012}
J. Dalibard, F. Gerbier, G. Juzeli\ifmmode~\bar{u}\else \={u}\fi{}nas, and P.
  \"Ohberg, Rev. Mod. Phys. {\bf 83},  1523  (2011).

\bibitem{Ho:2010a}
T.-L. Ho and Q. Zhou, Nature Physics {\bf 6},  131  (2010).

\bibitem{Nascimbene:2010}
S. Nascimb\`ene {\it et~al.}, Nature {\bf {463}},  1057  ({2010}).

\bibitem{Navon:2010}
N. Navon, S. Nascimb\`ene, F. Chevy, and C. Salomon, Science {\bf 328},  729
  (2010).

\bibitem{Hung:2011}
C.-L. Hung, X. Zhang, N. Gemelke, and C. Chin, Nature {\bf 470},  236  (2011).

\bibitem{Yefsah:2011}
T. Yefsah {\it et~al.}, Phys. Rev. Lett. {\bf 107},  130401  (2011).

\bibitem{Ku:2012}
M.~J.~H. Ku, A.~T. Sommer, L.~W. Cheuk, and M.~W. Zwierlein, Science {\bf 335},
   563  (2012).

\bibitem{Smith:2011b}
R.~P. Smith {\it et~al.}, Phys. Rev. Lett. {\bf 107},  190403  (2011).

\bibitem{Drake:2012}
T.~E. Drake {\it et~al.}, Phys. Rev. A {\bf 86},  031601  (2012).

\bibitem{Sagi:2012}
Y. Sagi, T.~E. Drake, R. Paudel, and D.~S. Jin, Phys. Rev. Lett. {\bf 109},
  220402  (2012).

\bibitem{Meyrath:2005}
T.~P. Meyrath {\it et~al.}, Phys. Rev. A {\bf 71},  041604  (2005).

\bibitem{Gupta:2005}
S. Gupta {\it et~al.}, Phys. Rev. Lett. {\bf 95},  143201  (2005).

\bibitem{Gotlibovych:2013}
I. {Gotlibovych} {\it et~al.}, arXiv:1212.4108  (2012).

\bibitem{Lin:2009}
Y.-J. Lin {\it et~al.}, Phys. Rev. A {\bf 79},  063631  (2009).

\bibitem{Kaplan:2002}
A. Kaplan, N. Friedman, and N. Davidson, J. Opt. Soc. Am. B {\bf 19},  1233
  (2002).

\bibitem{Jaouadi:2010}
A. Jaouadi {\it et~al.}, Phys. Rev. A {\bf 82},  023613  (2010).

\bibitem{Liesener:2000}
J. Liesener, M. Reicherter, T. Haist, and H. Tiziani, Optics Communications
  {\bf 185},  77   (2000).

\bibitem{Reinaudi:2007}
G. Reinaudi, T. Lahaye, Z. Wang, and D. Gu\'ery-Odelin, Opt. Lett. {\bf 32},
  3143  (2007).

\bibitem{Grossmann:1995}
S. Grossmann and M. Holthaus, Zeitschrift f\"ur Physik B Condensed Matter {\bf
  97},  319  (1995).

\bibitem{Arnold:2001}
P. Arnold and G. Moore, Phys. Rev. Lett. {\bf 87},  120401  (2001).

\bibitem{Kashurnikov:2001}
V.~A. Kashurnikov, N.~V. Prokof'ev, and B.~V. Svistunov, Phys. Rev. Lett. {\bf
  87},  120402  (2001).

\bibitem{Smith:2011}
R.~P. Smith, R.~L.~D. Campbell, N. Tammuz, and Z. Hadzibabic, Phys. Rev. Lett.
  {\bf 106},  250403  (2011).

\bibitem{Andrews:1997a}
M.~R. Andrews {\it et~al.}, Science {\bf 275},  637  (1997).

\bibitem{Navon:2011}
N. Navon {\it et~al.}, Phys. Rev. Lett. {\bf 107},  135301  (2011).

\bibitem{Wild:2012}
R.~J. Wild {\it et~al.}, Phys. Rev. Lett. {\bf 108},  145305  (2012).

\bibitem{Li:2012}
W. Li and T.-L. Ho, Phys. Rev. Lett. {\bf 108},  195301  (2012).

\bibitem{Gaunt:2012}
A.~L. Gaunt and Z. Hadzibabic, Sci. Rep. {\bf 2},  721  (2012).

\end{thebibliography}

\end{document}